\documentstyle[twoside,fleqn,espcrc2,epsf]{article}

\def\siml{{\ \lower-1.2pt\vbox{\hbox{\rlap{$<$}\lower6pt\vbox{\hbox{$\sim$}}}}\ }}

\def\als{\alpha_{\rm s}}

\def\lQ{\Lambda_{\rm QCD}}

\newcommand{\AmS}{{\protect\the\textfont2
  A\kern-.1667em\lower.5ex\hbox{M}\kern-.125emS}}

\newcommand{\be}{\begin{equation}}
\newcommand{\ee}{\end{equation}}
\newcommand{\bea}{\begin{eqnarray}}
\newcommand{\eea}{\end{eqnarray}}

\def\lQ{\Lambda_{\rm QCD}}

\def\als{\alpha_{\rm s}}
\def\siml{{\ \lower-1.2pt\vbox{\hbox{\rlap{$<$}\lower6pt\vbox{\hbox{$\sim$}}}}\ }} 
\def\simp{{\ \lower-1.2pt\vbox{\hbox{\rlap{$>$}\lower6pt\vbox{\hbox{$\sim$}}}}\ }}

\newcommand{\Appendix}[1]%
    {%
     \section{#1}%

      }

\title{Applications of Nonrelativistic Effective Field Theories to quarkonium systems with a small 
       radius}

\author{Nora Brambilla\address[MCSD]{INFN and Dipartimento di Fisica dell'universita' di Milano,\\ 
        Via Celoria 16, 20133 Milan, Italy}}

\begin{document}

\begin{abstract}
We review  some  predictions of nonrelativistic effective field theories for heavy quark-antiquark systems with a 
small typical radius, $r < 1/\Lambda_{\rm QCD}.$
\vspace{1pc}
\end{abstract}

\maketitle

\section{INTRODUCTION}
We are  interested here  in
bound states composed 
only by heavy quarks (and gluons). In such cases, at least for the lowest 
states, the characteristic radius $r$ ($r$ being the 
$q\bar{q}$ distance) of the system is  small, typically too small 
to probe the confinement effects, which arise at a scale ${1/\lQ}
 \gg r $. In the following  I will call 
such systems 'Coulombic' or `quasi-Coulombic'.
These  situations are particularly interesting 
not only because they have phenomenological 
relevance, but especially because they allow us to understand much more about QCD.
In such cases both the mass scale $m$ and the soft scale $1/r$ are much bigger 
than $\lQ$ and thus still sit in the perturbative regime.
Non-perturbative corrections exist 
but are  not expected to be dominant.\par
As it is apparent from the spectra, heavy quarkonia are non-relativistic systems.
Thus, they may  be described 
in first approximation using a Schr\"odinger equation with a potential 
interaction. This amounts to saying  that the heavy quark bound state is characterized
 by three energy scales, hierarchically ordered by the quark velocity 
$v \ll 1$: the quark mass $m$ (hard scale), 
the momentum $mv\simeq 1/r$ (soft scale (S)),  and 
the binding energy $mv^2$ (ultrasoft scale (US)). In the Coulombic or quasi-Coulombic
situation it is $v \sim \alpha_{\rm s}$.\par
To address the multiscale dynamics of the heavy quark bound state, 
the concept of effective field theory  turns out to be  not only helpful  but actually
necessary.
QCD effective field theories 
(EFT) with less and less degrees of freedom, can be obtained 
by systematically integrating out the scales  above the energy we aim to describe. 
This procedure leads ultimately to a  field theory 
derived quantum mechanical description  of these systems. The corresponding EFT is called 
pNRQCD \cite{pnrqcd}. Here, all the dynamical regimes
 are organized in a systematic expansion  in the ratio of the different scales which is 
eventually an expansion in $v$.  \par
In practice, by integrating out the hard scale NRQCD is obtained from QCD.
After integrating out the soft scale  in NRQCD, pNRQCD  is obtained. The Lagrangian of 
pNRQCD is organized in powers of $1/m$ and ${\bf r}$ (multipole expansion). 
The matching is  done by comparing appropriate off-shell amplitudes in NRQCD and in pNRQCD,
order by order in $1/m$, $\alpha_s$ and order by order in the multipole expansion.
The matching 
coefficients are non-analytic functions of ${\bf r}$  and have 
typically the following structure:
$ V \simeq {\cal V} ({\bf r}, {\bf p}, {\bf S_1}, {\bf S}_2) (A \ln mr +B
\ln \mu^\prime r +C)$.
\vskip -0.2truecm

\begin{table*}[htb]
\caption{Summary of the different kinematical situations.}
\label{tab1}
\makebox[0cm]{\phantom b}
\begin{tabular}{@{}llll}
\hline
$mv$&$mv^2$&potential&ultrasoft corrections\\\hline
$\gg \Lambda_{\rm QCD}$&$\gg \Lambda_{\rm QCD}$&perturbative&local condensates\\
$\gg \Lambda_{\rm QCD}$&$\sim \Lambda_{\rm QCD}$&perturbative&non-local condensates\\
$\gg \Lambda_{\rm QCD}$&$\ll \Lambda_{\rm QCD}$&perturbative + & No US (if light quarks\\
&$~$&short-range nonpert.& $\,\,\,$ are not considered)\\ 
\hline
\end{tabular}\\[2pt]
\end{table*}

\section{pNRQCD for $mv \gg \lQ$}

 At the scale of the matching $\mu^\prime$ 
($mv \gg \mu^\prime \gg mv^2,  \lQ$) we have still quarks and gluons.
We denote by ${\bf R}\equiv {({\bf x}_1+{\bf x}_2})/2$ the center-of-mass of the $q\bar{q}$ 
system and 
by  ${\bf r\equiv {\bf x}_1 -{\bf x}_2}$ the relative distance. 
The effective degrees of freedom are: $q\bar{q}$ states (that can be decomposed into 
a singlet $S({\bf R},{\bf r},t)$ and an octet $O({\bf R},{\bf r},t)$
under color transformations) with energy of the order of the next relevant 
scale, $O(\Lambda_{QCD},mv^2)$, and momentum\footnote{Although,
for simplicity, we describe the matching between NRQCD and pNRQCD as integrating out 
the soft scale, the relative momentum ${\bf p}$ of the quarks is
still soft.}  ${\bf p}$ of order $O(mv)$,  plus 
ultrasoft gluons $A_\mu({\bf R},t)$ with energy 
and momentum of order  $O(\lQ,mv^2)$. Notice that {\it all the  gluon fields}
 are {\it multipole 
expanded}. The Lagrangian is then  an expansion 
in the small quantities  $ {p/m}$, ${ 1/r  m}$ and in   
$O(\Lambda_{\rm QCD}, m v^2)\times r$.
The pNRQCD Lagrangian is given in this case
at the leading order in the multipole expansion by\cite{pnrqcd}:
\begin{eqnarray}
& &\hspace{-4mm}
{\cal L}=
{\rm Tr} \Biggl\{ {\rm S}^\dagger \left( i\partial_0 - {{\bf p}^2\over m} 
- V_s(r) -\sum_{n=1} {V_s^{(n)}\over m^n} \right) {\rm S} 
\nonumber \\
& & \!\!\!\!\!\!\!\!\!\!+ {\rm O}^\dagger \left( iD_0 - {{\bf p}^2\over m} 
- V_o(r) - \sum_{n=1} {V_o^{(n)}\over m^n}  \right) {\rm O} \Biggr\}
\nonumber\\
& & \!\!\!\!\!\!\!\!\!\!
+ g V_A ( r) {\rm Tr} \left\{  {\rm O}^\dagger {\bf r} \cdot {\bf E} \,{\rm S}
+ {\rm S}^\dagger {\bf r} \cdot {\bf E} \,{\rm O} \right\} 
\label{pnrqcd0}\\
& &\!\!\!\!\!\!\!\!\!\!\!
   + g {V_B (r) \over 2} {\rm Tr} \left\{  {\rm O}^\dagger {\bf r} \cdot {\bf E} \, {\rm O} 
+ {\rm O}^\dagger {\rm O} {\bf r} \cdot {\bf E}  \right\} -{1\over 4} F^a_{\mu\nu}
F^{\mu \nu a}.  
\nonumber
\end{eqnarray}

All the gauge fields in Eq. (\ref{pnrqcd0}) are evaluated 
in ${\bf R}$ and $t$. In particular ${\bf E} \equiv {\bf E}({\bf R},t)$ and 
$iD_0 {\rm O} \equiv i \partial_0 {\rm O} - g [A_0({\bf R},t),{\rm O}]$. 
The quantities denoted by $V$ are the matching coefficients.

We call $V_s$ and $V_o$ the singlet and octet static matching potentials respectively.
At the leading order in the multipole expansion,
the singlet sector of the Lagrangian gives rise to equations of motion of the 
Schr\"odinger type. The two last lines of (\ref{pnrqcd0})
contain (apart from the Yang-Mills Lagrangian) retardation (or non-potential) effects that 
start at the NLO in the multipole expansion. At this order the non-potential
effects come from the singlet-octet and octet-octet 
interactions mediated by a ultrasoft chromoelectric 
field. 
Recalling that  
 ${\bf r} \sim 1/mv$ and that the operators count like the next relevant 
scale, $O(mv^2,\lQ)$, to the power of the dimension, it follows that  
each term in  the pNRQCD Lagrangian has a definite power counting.  This feature makes 
${\cal L}_{\rm pNRQCD}$ a suitable tool for  bound state calculations: being interested 
in knowing the energy levels up to some power $v^n$, we just need to 
evaluate the contributions
of this size from the Lagrangian. From the power 
counting e.g.,  it follows that the interaction of quarks with ultrasoft 
gluons (nonpotential or retardation effect) is suppressed in the Lagrangian
 by $v$ with respect to the LO ( by $g v$ if $mv^2 \gg \lQ$).

In particular, pNRQCD provides us with the way of obtaining the matching 
potentials via 
the matching procedure at any order of the perturbative expansion\cite{pnrqcd}.
In the EFT language the potential is defined upon 
the integration of all the scales {\it up to the ultrasoft 
scale $mv^2$}.  From the matching to NRQCD  in the situation $\lQ \ll mv$
we can easily obtain the  matching potential $V_s$ 
at ${\rm N}^3$LL \cite{pnrqcd}
\begin{eqnarray}
& & \hspace{-7mm} 
V_s(r) \equiv  - C_F {\alpha_{V}(r,\mu^\prime) \over r}, \nonumber\\
& & \hspace{-7mm} 
{\alpha}_{V}(r, \mu)=\alpha_{\rm s}(r)
\left\{1+\left(a_1+ 2 {\gamma_E \beta_0}\right) {\alpha_{\rm s}(r) \over 4\pi}\right. \nonumber\\
& &\hspace{-5mm} 
+{\alpha_{\rm s}^2(r) \over 16\,\pi^2}
\bigg[\gamma_E\left(4 a_1\beta_0+ 2{\beta_1}\right)+\left( {\pi^2 \over 3}+4 \gamma_E^2\right) 
{\beta_0^2}
\nonumber\\
& &\hspace{12mm} 
+ a_2\bigg] \left. + {C_A^3 \over 12}{\alpha_{\rm s}^3(r) \over \pi} \ln{ r \mu^\prime}\right\},
\label{newpot}
\end{eqnarray}
where $\beta_n$ are the coefficients of the beta function ($\alpha_{\rm s}$ is in the $\overline{\rm MS}$ scheme), 
and $a_1$ and $a_2$ were given in \cite{pert}. 
The Coulomb potential turns out to be sensitive to the ultrasoft scale but infrared finite. The same 
happens with the other potentials (like $V_o$ or the potentials that bear corrections in $1/m^n$)
 that can equally be calculated via the matching procedure.
The $\mu^\prime$ scale dependence in the potential is cancelled for any physical 
process by the contribution of the ultrasoft gluons that are cutoff at the scale 
$\mu^\prime$.\par  Then, there are two main situations \cite{pnrqcd}, cf. Table 1.
 If
$mv\gg mv^2\simp \lQ$, 
the system is described up to order $\alpha_{\rm s}^4$ by a potential which 
is entirely accessible to perturbative QCD. 
Non-potential effects start at order $\alpha_{\rm s}^5\ln \mu^\prime$ \cite{nnll,onnll}. 
We call {\it Coulombic}
 this kind of system. Non-perturbative effects are of non-potential type
and can be encoded into local 
(\`a la Voloshin--Leutwyler) or non-local condensates:
 they are suppressed by powers of  $\lQ/mv^2$ and $\lQ/mv$ respectively. 
If $mv \gg \lQ \gg m v^2$, 
 the scale $mv$ can be still integrated out perturbatively,
giving rise to the Coulomb-type  potential (\ref{newpot}).
Non-perturbative  contributions to the potential arise 
when integrating out the scale $\lQ$ \cite{pnrqcd}.
We call {\it quasi-Coulombic} the  systems  where 
the non-perturbative piece of the potential 
can be considered small with respect to the Coulombic one and treated as a perturbation.
Some levels of $t\bar{t}$, the lowest level of $b \bar{b}$  may be considered Coulombic systems,
while the $J/\psi$, the $\eta_c$ and the short-range hybrids may be considered quasi-Coulombic.
The   $B_c$  may be in a boundary situation.
As it is typical in an effective theory,
 only the actual calculation may confirm if the initial
assumption about the physical system was appropriate.

\begin{table*}[htb]
\caption{\footnotesize\it
The quantities $E_X^{(j)}$ are the order $\varepsilon^j$ contributions to the spectrum
\cite{sumino,epsilon}.
The $c$-  and $b$-quark ${\rm msbar}$ masses are fixed on the experimental 
values of the $J/\psi$ and $\Upsilon(1S)$ 
masses.
The uncertainties in the third and fourth columns refer 
to the uncertainties in $\als^{(5)}(M_Z)$ only. All the other data refer to
$\als^{(5)}(M_Z)=0.1181$ and to the ${\rm msbar}$ quark 
masses fixed on the central values.}
\begin{tabular}{@{}lllllllll}
\hline
State $X$ &$E_X^{\rm exp}~~$ &$E_X~~~$ & $E_X^{\rm exp}-E_X$ 
&$E_X^{(1)}$ &$E_X^{(2)}$ &$E_X^{(3)}$ &~~$\mu_X^A$ &$\als^{(3)}(\mu_X^A)$ \\ 
\hline
$\Upsilon(1^3S_1)$  & 9.460 & 9.460~~~~ & 0~~~~~ & 0.866 & 0.208  & 0.006 & 2.14 & 0.286 \\
$\Upsilon(1 ^3P_0)$ & 9.860 & $9.995^{+75}_{-62}$ & $-$0.135$^{-75}_{+62}$ & 1.534  & 0.101  &$-$0.021 & 1.08 & 0.459 \\
$\Upsilon(1 ^3P_1)$ & 9.893 &$10.004^{+78}_{-63}$ & $-$0.111$^{-78}_{+63}$ & 1.564  & 0.081  &$-$0.022 & 1.05 & 0.468 \\
$\Upsilon(1 ^3P_2)$ & 9.913 &$10.012^{+81}_{-65}$ & $-$0.099$^{-81}_{+65}$  & 1.591  & 0.063  &$-$0.022 & 1.034 & 0.477 \\
$\Upsilon(2 ^3S_1)$ &10.023 &$10.084^{+93}_{-75}$ & $-$0.061$^{-93}_{+75}$ & 1.618  & 0.096  &$-$0.010 & 1.02 & 0.486 \\
$\Upsilon(2 ^3P_0)$ &10.232 &$10.548^{+196}_{-151}$ & $-$0.316$^{-196}_{+151}$  & 2.421  &$-$0.356  & 0.102 & 0.778 & 0.710 \\
$\Upsilon(2 ^3P_1)$ &10.255 &$10.564^{+200}_{-153}$ & $-$0.309$^{-200}_{+153}$ & 2.472  &$-$0.404  & 0.116 & 0.770 & 0.726 \\
$\Upsilon(2 ^3P_2)$ &10.268 &$10.578^{+203}_{-155}$ & $-$0.310$^{-203}_{+155}$ & 2.518  &$-$0.449  & 0.129 & 0.762 & 0.740 \\
$\Upsilon(3 ^3S_1)$ &10.355 &$10.645^{+218}_{-168}$ & $-$0.290$^{-218}_{+168}$  & 2.472  &$-$0.348  & 0.140 & 0.770 & 0.726 \\
$\Upsilon(4 ^3S_1)$ &10.580 &$*$                    & $*$ & $*$  &$*$ & $*$ & $*$ & $*$ \\
$B_c(1^1S_0)$ & $6.4 \pm 0.4$ & $6.307^{+4}_{-2}$ & 0.1$\pm 0.4$ & 0.675 & 0.188 & 0.017 & 1.62 & 0.334 \\
\hline
\vspace{1mm}
\end{tabular}\\
\end{table*}

For all these systems it is relevant to obtain a determination of the energy levels 
as accurate as possible in perturbation theory.
In recent years our knowledge of heavy quarkonia in the framework of perturbative QCD 
has developed significantly. On one hand computations of new higher-order corrections 
to various physical quantities appeared e.g.\cite{nnll,onnll,mass}. 
On the other hand the discovery 
of the mechanism of the renormalon cancellation in the quarkonium spectrum 
led to a drastic improvement of the convergence of the perturbative 
expansion of the energy levels.  As important applications up to date, the theory enabled precise 
determinations of the $\overline{\rm MS}$-mass of the bottom quark 
and of the top quark  from (mainly) the energy spectra 
of the lowest-lying states. The main uncertainty comes, in the bottomonium case, 
from the (essentially) unknown non-perturbative contributions. These are generally claimed 
to be around 100 MeV and ultimately set the precision of the prediction. 
Within pNRQCD 
a complete and systematic parametrization  of perturbative and non-perturbative effects 
of the heavy quarkonium interaction can be performed (cf. Table 1 and \cite{pnrqcd}).
The parametrization depends on the mutual relation between the scale of non-perturbative physics, $\lQ$, 
and the other dynamical scales in the specific heavy quarkonium system under study.
A way to determine it and the size and nature of the non-perturbative contributions, 
consists in establishing to which extent perturbative QCD can consistently and successfully 
describe the system. 
This is the problem we have addressed in\cite{sumino}, which investigates the range of validity 
of perturbative QCD on the heavy quarkonium spectroscopy and consequently extracts upper bounds 
to the non-perturbative contributions by comparing the perturbative predictions, 
at the current best accuracy, with the experimental data.
One of the main problems in having a consistent (i.e. convergent) perturbative 
expansion for the quarkonium energy levels has been for a long time its bad behaviour 
when expressed in terms of the pole mass. 
The reason has been understood in the renormalon 
language\cite{renormalon}: the pole mass and the QCD static potential, respectively, 
contain renormalon contributions of order $\lQ$, which get cancelled in the total energy of a color singlet
quark-antiquark system.
The solution then consists in making explicit this cancellation by substituting in the energy expansion  
a short-range mass (free from infrared ambiguities) for the pole mass.
In order to make explicit the renormalon cancellation 
we will express the quarkonium energies in terms of the $\overline{\hbox{MS}}$ masses 
and rearrange the perturbative series in the so-called $\varepsilon$ expansion \cite{epsilon}. 
This will be the key ingredient of our analysis. We note that according to a formal argument, 
the residual uncertainty of the perturbative expansion due to the next-to-leading renormalon 
contribution is estimated to be  $\Lambda_{\rm QCD}\times (a_X \Lambda_{\rm QCD})^2$ for a bound-state 
$X$ of size $a_X$ \cite{pnrqcd,sumino}.
Provided this argument applies, in principle the predictions of perturbative QCD can be made 
precise down to this accuracy by sufficiently increasing the orders of the perturbative expansion.
The complete analysis of the consistency 
between the whole level structure of perturbative QCD and the experimental data has been done in \cite{sumino}. 
The major results of that analysis have been as follows (see Table 2).
(1) Once the cancellation of the leading renormalon has been incorporated,
the perturbative series turns out to be convergent and to reproduce reasonably well 
the gross structure of the bottomonium spectrum at least up to some of the 
$n=3$ levels. (2) The constraints on non-perturbative and higher-order contributions 
to the bottomonium spectrum, set by the comparison of the calculation with 
the experimental data, indicate that these are smaller than usually believed.
Moreover, the perturbative predictions of the bottomonium spectrum up to $n=3$ appears to 
be in agreement with the experimental data (within  the next-to-leading renormalon uncertainties and taken into account the error with which we know $\alpha_s$).


\vskip -1truecm
\begin{thebibliography}{9}
\bibitem{pnrqcd}
N.~Brambilla, A.~Pineda, J.~Soto and A.~Vairo,
Nucl.\ Phys.\ B {\bf 566}, 275 (2000)
N.~Brambilla, A.~Pineda, J.~Soto and A.~Vairo,
Phys.\ Rev.\ D {\bf 60}, 091502 (1999)
A.~Pineda and J.~Soto,
Nucl.\ Phys.\ Proc.\ Suppl.\  {\bf 64}, 428 (1998)
\bibitem{pert}
Y.~Schroder,
Phys.\ Lett.\ B {\bf 447}, 321 (1999)
M.~Peter,
Nucl.\ Phys.\ B {\bf 501}, 471 (1997)
\bibitem{nnll}
N.~Brambilla, A.~Pineda, J.~Soto and A.~Vairo,
Phys.\ Lett.\ B {\bf 470}, 215 (1999)
\bibitem{onnll}
B.~A.~Kniehl, A.~A.~Penin, V.~A.~Smirnov and M.~Steinhauser,
Nucl.\ Phys.\ B {\bf 635}, 357 (2002)
\bibitem{sumino} 
N.~Brambilla, Y.~Sumino and A.~Vairo,
Phys.\ Rev.\ D {\bf 65}, 034001 (2002)
N.~Brambilla, Y.~Sumino and A.~Vairo,
Phys.\ Lett.\ B {\bf 513}, 381 (2001)
S.~Recksiegel and Y.~Sumino,
S.~Recksiegel and Y.~Sumino,
arXiv:hep-ph/0207005.
N.~Brambilla and A.~Vairo,
Phys.\ Rev.\ D {\bf 62}, 094019 (2000)
[arXiv:hep-ph/0002075].
\bibitem{mass}
A.~A.~Penin and M.~Steinhauser,
Phys.\ Lett.\ B {\bf 538}, 335 (2002)
A.~Pineda,
JHEP {\bf 0106}, 022 (2001)
J.~H.~Kuhn and M.~Steinhauser,
Nucl.\ Phys.\ B {\bf 619}, 588 (2001)
[Erratum-ibid.\ B {\bf 640}, 415 (2002)]
A.~H.~Hoang,
\bibitem{renormalon}
A.~H.~Hoang, M.~C.~Smith, T.~Stelzer and S.~Willenbrock,
Phys.\ Rev.\ D {\bf 59}, 114014 (1999)
\bibitem{epsilon}
A.~H.~Hoang, Z.~Ligeti and A.~V.~Manohar,
Phys.\ Rev.\ Lett.\  {\bf 82}, 277 (1999)
\end{thebibliography}
\end{document}